\documentclass[aps
,preprint
,groupedaddress
]{revtex4-1}
\usepackage{graphicx}
\usepackage[misc,geometry]{ifsym}
\usepackage{amsmath}
\usepackage{color}

\usepackage{fancyhdr}
\usepackage[bookmarksnumbered, bookmarksopen=true, colorlinks, citecolor=blue, linkcolor=blue]{hyperref}

\DeclareMathOperator{\f}{\textbf{\emph{f}}}

\DeclareMathOperator{\x}{\textbf{\emph{x}}}
\DeclareMathOperator{\B}{\textbf{\emph{B}}}
\DeclareMathOperator{\D}{\textbf{\emph{D}}}
\DeclareMathOperator{\W}{\textbf{\emph{W}}}
\DeclareMathOperator{\R}{\textbf{\emph{R}}}
\DeclareMathOperator{\Q}{\textbf{\emph{Q}}}
\DeclareMathOperator{\T}{\textbf{\emph{T}}}
\DeclareMathOperator{\X}{\textbf{\emph{X}}}
\renewcommand\S{\textbf{\emph{S}}}
\renewcommand\d{\mathrm{d}}
\renewcommand\r{}

\newtheorem{thm:thm}{Theorem}[section]
\newtheorem{prf:prf}{Proof}[section]
\newtheorem{ex:ex}{Example}[section]

\begin{document}
\title{Relation of a New Interpretation of Stochastic Differential Equations to Ito Process}
\author{Jianghong Shi}
\email[ Email: ]{zjjxsjh@gmail.com}
\affiliation{Department of Computer Science and Engineering, Shanghai Jiao Tong University, Shanghai, 200240, China.}
\author{Tianqi Chen}
\email[ Email: ]{tqchen@apex.sjtu.edu.cn}
\affiliation{Department of Computer Science and Engineering, Shanghai Jiao Tong University, Shanghai, 200240, China.}
\author{Ruoshi Yuan}
\email[ Email: ]{yrs12345@sjtu.edu.cn}
\affiliation{Department of Computer Science and Engineering, Shanghai Jiao Tong University, Shanghai, 200240, China.}
\author{Bo Yuan}
\email[Corresponding author. Email: ]{boyuan@sjtu.edu.cn}
\affiliation{Department of Computer Science and Engineering, Shanghai Jiao Tong University, Shanghai, 200240, China.}
\author{Ping Ao}
\email[Corresponding author. Email: ]{aoping@sjtu.edu.cn}
\affiliation{Shanghai Center for Systems Biomedicine, Key Laboratory of Systems Biomedicine of Ministry of Education, Shanghai Jiao Tong University, Shanghai, 200240, China.\\
Institute of Theoretical Physics, Shanghai Jiao Tong University, Shanghai, 200240, China.}

\begin{abstract}
Stochastic differential equations (SDE) are widely used in modeling stochastic dynamics in literature. However, SDE alone is not enough to determine a unique process. A specified interpretation for stochastic integration is needed. Different interpretations specify different dynamics. Recently, a new interpretation of SDE is put forward by one of us. This interpretation has a built-in Boltzmann-Gibbs distribution and shows the existence of potential function for general processes, which reveals both local and global dynamics. Despite its powerful property, its relation with classical ones in arbitrary dimension remains obscure. In this paper, we will clarify such connection and derive the concise relation between the new interpretation and Ito process. We point out that the derived relation is experimentally testable.
\end{abstract}
\maketitle

\section{Introduction}\label{intro}
Stochastic processes universally exist in nature as well as human society. One of the most popular tools, stochastic differential equation (SDE) has a long history of being used to model continuous processes in physics and chemistry \cite{gardiner,kampen}. Recently, there is a growing trend to apply it in biology, from complex networks \cite{ao_cancer_2010,ao_cancer_2008,qian_2010,wang_2008} to evolution dynamics \cite{jiao_2011,zhou_2011}.

However, due to the intrinsic richness of SDE, one will encounter an ambiguity in choosing integration method when applying SDE to model practical problems \cite{volpe_2010}. There are already some well established integration methods in literature, Ito, Stratonovich, for example \cite{gardiner}. It is known that they are consistent in their own mathematical framework and their relations have been addressed \cite{lyons_1998,ilya_2010,wong_1965}. \r{While researchers are aware that different interpretations of SDE have different applied meanings \cite{sussman_1978}, it turns out that in practice} people tend to choose integration method by professional preference: mathematician prefer Ito due to martingale property, engineers prefer Stratonovich owing to chain rule. What we want to emphasize here is that when dealing with practical problems, careful consideration should be taken for such ambiguity.

The fact that different integration methods lead to different processes can be best shown by their corresponding Fokker-Planck equations (FPE). In contrast to SDE, FPE is complete and deterministic, describing the ensemble evolution of processes. A given SDE with a specified integration method corresponds to a unique FPE. Accordingly, A given SDE with a specified FPE indicates an integration method. Thus, to determine a system described by an SDE, one can either specify the integration method, or assign a certain FPE to it.

Recently, a new interpretation framework for SDE is formed by one of us while studying a biological stability problem \cite{ao_phage_2004}.  This interpretation, instead of specifying the integration method of SDE explicitly, transforms the SDE into a canonical form (Equation \ref{eq:ao sde}) and links it with a special FPE with the same structure. This novel construction shows the existence of the potential function \cite{mu_2011,yuan_2011} for general processes, which reveals local dynamics via SDE and global dynamics via Boltzmann-Gibbs distribution.

Real-world systems using this kind of SDE interpretation have been found \cite{volpe_2010}. Related issues have been considered along the line \cite{ao_2008,ao_2007,ao_2005,ao_2011,ao_2006}. However, the relation between the new interpretation and classical ones in arbitrary dimension has not been clearly addressed. In addition, such relation may have experimental implications. In this paper, we will derive the relation between the new interpretation and Ito process and discuss how to tell the differences between different interpretations.

We will review classical SDE integration methods in section 2. The new interpretation will be introduced in section 3, followed by its relation with Ito process. Conclusion can be found in the end.
\section{Classical SDE integration methods}
\label{sec:1}
In this section, we review the definitions of classical SDE integration methods and derive the relation between them. There are usually two ways to write an SDE, namely the Langevin form in physics and the mathematical form \cite{kampen}. We apply the latter in this paper.

A typical mathematical formulation of SDE is given by the following formula
\begin{equation}\label{eq:sde}
\d \x = \f(\x(*)) \d t + \B(\x(*)) \d\W (t), \B\B ^{\tau} = 2\theta \D(\x).
\end{equation}
For simplicity, we only consider time homogeneous process. Here $\x$ is a $n$ dimensional variable and $\f(\x)$ is the drift force, either linear or non-linear. $\d\W(t)$ is a $m$ dimensional Gaussian white noise with zero mean and variance $\d t$. $\B(\x)$ is a $n\times m$ matrix. $\theta$ is a non-negative constant playing the role of temperature and $\D(\x)$ is the symmetric and positive semi-definite diffusion matrix. SDE does not contain information for choosing $\x(*)$ here, that is where different integration methods differ from.
In the following, we will talk about the connections between classical interpretations. Although this is known, such demonstration will make our later discussion more clear.

Consider a time interval $[t, t+\d t]$, a general form which we call $\alpha$-type integration interprets Equation \ref{eq:sde} as follows
\begin{equation}\label{eq:alpha sde}
\d \x = \f(\x(t+\alpha \d t)) \d t +\B(\x(t+\alpha \d t)) \d \W(t), \B\B^\tau = 2\theta \D(\x),\alpha \in [0,1].
\end{equation}
To see why $\alpha$ matters here, we shall take a close look at the approximate orders. In fact, like ordinary differential equation, $\alpha$ does not affect the deterministic term. This is because different ways of choosing $\alpha$ for the drift force only lead to a difference of order $\mathrm{o}(\d t)$. But it is another story for SDE. Note that $\d \x$ is of order $\mathrm{O}(\sqrt{\d t})$ due to the noise $\d \W(t)$. This leads to an order $\mathrm{O}(\sqrt{\d t})$ change of $\B(\x)$, combined with $\d \W(t)$ resulting order $\mathrm{O}(\d t)$ change of $\d \x$. Thus, choosing different $\alpha$ would lead to a total difference of order $\mathrm{O}(\d t)$ for $\d \x$, which can not be ignored. This issue has been addressed mathematically by Wong and Zakai in 1965 \cite{wong_1965}. Noting that the Ito \cite{oksendal}, Stratonovich \cite{kampen}, and H\"{a}nggi \cite{hanggi_1978} integration methods correspond to $\alpha=0,\alpha=1/2,\alpha=1$ type processes separately.

\r{The following theorem summarizes the relation between $\alpha$-type and Ito type (I-type) integration method and an intuitive derivation can be found in appendix.}
\begin{thm:thm}\label{thm:ito alpha relation} ($\alpha$-I relation) \\
For an $\alpha$-type SDE Equation \ref{eq:alpha sde}, the equivalent I-type SDE is
\begin{equation}\begin{split}\label{eq:sde relation}
  \d \textbf{x} & = \left[\textbf{f}(\textbf{x})+ \textbf{h}(\textbf{x})\right] \d t + \textbf{B}(\textbf{x}) \d \textbf{W}(t) \\
\mathrm{where\;}  \textbf{h}_{i}(\textbf{x}) & = \alpha \sum_{j} \sum_{k} (\partial_{k} \textbf{B}_{ij}(\textbf{x})) \textbf{B}_{kj}(\textbf{x}).
\end{split}\end{equation}
\end{thm:thm}
With above results, one can use I-type process to simulate $\alpha$-type process with a modification of drift force. Note that for a specific SDE with constant $\B$, all $\alpha$-type interpretations lead to an identical process.

In the end of this section, we mention that in order to uniquely determine a process, one can specify the interpretation for Equation \ref{eq:sde}. Alternatively, one can also write the Fokker-Planck equation (FPE) of the process. The FPE describes the evolution of probability distribution function of a process. A given SDE with a specific interpretation describes a determined process, thus corresponds to a determined FPE. The FPE of Equation \ref{eq:sde} with I-type interpretation is \cite{gardiner,karatzas,oksendal,kampen}
\begin{equation}\label{eq:ito fpe}
\partial_{t} \rho(\x) = -\sum_{i}\partial_{i}[\f_{i}(\x)\rho(\x)]+ \theta\sum_{i,j}\partial_{i}\partial_{j}[\D_{ij}(\x)\rho(\x)].
\end{equation}
With the relation we obtained, one can get the corresponding $\alpha$-type FPE according to Equation \ref{eq:sde relation}.

\section{Potential function view interpretation of SDE}
In this section, we introduce the potential function view interpretation of SDE formed by one of us recently. We call this A-type for short. It first decomposes the deterministic force and rewrite Equation \ref{eq:sde} into the following form
\begin{equation}\label{eq:ao sde_o}
\d \x = -[\D(\x)+\Q(\x)]\nabla \phi(\x)\d t + \B(\x)*\d\W(t),  \B\B^\tau = 2\theta \D(\x),
\end{equation}
where $\D$ is the symmetric positive semi-definite diffusion matrix and $\Q$ is an anti-symmetric matrix. Here $\phi$ is a scaler function playing the role of potential function in physics. Then it transforms the above equation into the following equation
\begin{equation}\label{eq:ao sde}
[\S(\x)+\T(\x)]\d\x = -\nabla \phi(\x)\d t + \hat \B(\x)*\d\W(t), \hat \B\hat \B^\tau = 2\theta \S(\x),
\end{equation}
where $\S$ is a symmetric positive semi-definite matrix and $\T$ is an anti-symmetric matrix.
The transformation is based on the following assumption: Deterministic part and stochastic part corresponds to the original ones separately. Subsequently the relation, the so-called generalized Einstein relation \cite{ao_2008}, directly follows:
\begin{equation}
[\S(\x)+\T(\x)]\D(\x)[\S(\x)-\T(\x)]=\S(\x).
\end{equation}
\r{In one dimensional case, the above relation reduces to the well known Einstein relation. For higher dimensional cases, the generalized Einstein relation is a new prediction from our approach and can be tested experimentally. }

\r{We have a power series expansion method \cite{ao_2004} to generally construct the form (Equation \ref{eq:ao sde}). Furthermore, situation near fixed points are exhaustively studied \cite{ao_2005} and limit cycle case is explicitly constructed \cite{zhu_2006}. Equation \ref{eq:ao sde} is analogous to the the dynamics of a ``massless'' charged particle in viscous liquid with presence of electrical and magnetic field, where $\S$ and $\T$ play the role of dissipative force and magnetic field separately and $\nabla \phi(\x)$ as the electrical field, a detailed example can be found in \cite{ao_2008}. Note that the physical status of potential function $\phi(\x)$ is same as classical ones, with the capability of generalizing into non-equilibrium situation.}

Via applying a process of zero mass limit with $2n$ dimensional Klein-Kramers equation in physics, it has been shown that Equation \ref{eq:ao sde} corresponds to the following $n$ dimensional Fokker-Planck equation \cite{ao_2006}
\begin{equation}\label{eq:ao fpe}
\partial_{t}\rho(\x)=\nabla^\tau \{[\D(\x)+\Q(\x)][\nabla\phi(\x)+\theta\nabla]\rho(\x)\}.
\end{equation}
\r{To our best knowledge, previous works concerning zero mass limit procedure such as \cite{freidlin_2004,volpe_jsp,kupferman_2004} have never reached our form of SDE. Our approach involves both magnetic field and Einstein relation. Note that this procedure is also consistent within A-type framework \cite{ao_2007}.}
Equation \ref{eq:ao fpe} has a very nice property that its stationary distribution is in the form of Boltzmann-Gibbs distribution
\begin{equation}\label{eq:bgd}
\pi(\x)\propto \exp \left[-\frac{\phi(\x)}{\theta}\right].
\end{equation}

This new construction shows that one can find a potential function with both local and global meanings for a general process, even in cases without detailed balance condition, i.e., with nonzero $\T$. The potential function is responsible for local dynamics via SDE and global state via Boltzmann-Gibbs distribution. It is quite powerful that with it one can find out the final state of the system without going through the dynamics. Apart from the potential function, $\S$ and $\T$ that are responsible for dynamics also has physical meanings, namely the dissipative force and the magnetic field. In addition, combining Equation \ref{eq:ao sde} and Equation \ref{eq:bgd}, we can see that the states with extremal stationary probability coincides with the states with zero drift force.

With those nice properties, this method has been successfully applied to solve an important biological stability problem \cite{ao_phage_2004}. Just as an engineering experiment show good agreement with Stratonovich integration method \cite{mcclintock_1983}, a recent physical experiment \cite{volpe_2010} turn out to favor the A-type integration in one dimensional case. \r{As generalized Einstein equation can be experimentally tested, we are looking forward to higher dimensional experimental results. }

Notice that this new construction does not explicitly specify a classical integration method for the original SDE. Instead, it specifies the dynamics by linking an FPE to the original SDE. One may wonder what is the relation between this construction and the classical ones and how to choose a specific interpretation to deal with real problems. We will address these questions in the following.

\subsection{A-I relation}
Since we know the FPE for all the different integration methods, we can seek their relations from FPE, that is, the connection based on distribution function perspective. The relation between A-type and I-type interpretations are given by the following theorem.
\begin{thm:thm} (A-I Relation)\\
A-type SDE in the following form
\begin{equation}
\d\textbf{x} = -[\textbf{D}(\textbf{x})+\textbf{Q}(\textbf{x})]\nabla \phi(\textbf{x})\d t + \textbf{B}(\textbf{x})*\d\textbf{W}(t)
\end{equation}
Can be converted into I-type SDE
\begin{equation}\begin{split}
\d\textbf{x} & = \textbf{f}(\textbf{x}) \d t + \textbf{B}(\textbf{x})\d\textbf{W}(t) \\
   & = \{-[\textbf{D}(\textbf{x})+\textbf{Q}(\textbf{x})]\nabla \phi(\textbf{x})+\theta\Delta \textbf{f}(\textbf{x})\} \d t +  \textbf{B}(\textbf{x})\d\textbf{W}(t),
\end{split}\end{equation}
where the additional drift term is defined by
\begin{equation}\label{eq:add}
\Delta \textbf{f}_{i}(\textbf{x}) =\sum_{j}\partial_{j}[\textbf{D}_{ij}(\textbf{x})+\textbf{Q}_{ij}(\textbf{x})].
\end{equation}
\end{thm:thm}

\begin{prf:prf}
\label{temp}
In order to convert A-type FPE into I-type FPE, we rewrite Equation \ref{eq:ao fpe} into a more detailed form
\begin{equation}\begin{split}
\partial_{t}\rho(\x) = & \sum_{i}\partial_{i}\left\{ \sum_{j}[\D_{ij}(\x)+\Q_{ij}(\x)]\rho(\x)\partial_{j}\phi(\x) \right\} \\
& + \theta\sum_{i}\partial_{i}\left\{ \sum_{j}[\D_{ij}(\x)+\Q_{ij}(\x)]\partial_{j}\rho(\x)\right\}.
\end{split}\end{equation}
We want to convert it into I-type FPE (Equation \ref{eq:ito fpe}). By comparing the two equations, we can find the difference lies in the second part of both equations. We start from decomposing the second part of I-type FPE
\begin{equation}\begin{split}
  &\theta\sum_{i,j}\partial_{i}\partial_{j}[\D_{ij}(\x)\rho(\x)] \\
= &\theta\sum_{i,j}\partial_{i}\{[\partial_{j}\D_{ij}(\x)]\rho(\x)\} + \theta\sum_{i,j}\partial_{i}\{\D_{ij}(\x)\partial_{j}\rho(\x)\}.
\end{split}\end{equation}
We can find the second term is part of the second term in Equation \ref{eq:ao fpe}. For the remaining part in Equation \ref{eq:ao fpe}, we analyze it further as
\begin{equation}\label{eq:proof}\begin{split}
  &\theta\sum_{i,j}\partial_{i}[\Q_{ij}(\x)\partial_{j}\rho(\x)] \\
= & \theta\sum_{i,j}\left\{\partial_{i}\partial_{j}\left[\Q_{ij}(\x)\rho(\x)\right]-\partial_{i}\rho(\x)\partial_{j}\Q_{ij}(\x)\right\}\\
= & -\theta\sum_{i}\partial_{i}\left\{\left[ \sum_{j}\partial_{j}\Q_{ij}(\x) \right] \rho(\x) \right\} \; \;\;\mathrm{(by\: anti-symmetric\: property\: of}\: \Q(\x)) .
\end{split}\end{equation}
With the above results, we are ready to transform the second term of A-type FPE
\begin{equation}\begin{split}
&\theta\sum_{i}\partial_{i}\left\{ \sum_{j}[\D_{ij}(\x)+\Q_{ij}(\x)]\partial_{j}\rho(\x) \right\} \\
= & -\theta\sum_{i}\partial_{i}\left\{\left[ \sum_{j}\partial_{j}(\D_{ij}(\x)+\Q_{ij}(\x)) \right]\rho(\x) \right\} + \theta\sum_{i,j}\partial_{i}\partial_{j}[\D_{ij}(\x)\rho(\x)],
\end{split}\end{equation}
noting that the first term of A-type FPE is already same with I-type, we can find that the additional drift term is as Equation \ref{eq:add} shows.
\end{prf:prf}

Although A-type framework implicitly, via $2n$ dimensional Klein-Kramers equation, determines the integration method by linking an FPE to the given SDE in $n$ dimension, we may simulate its process with modified I-type process using the above relation directly in $n$ dimension. We should also note the following fact: In one dimensional case where $\Q$ vanishes, the additional fixing term makes A-type FPE to be $\alpha$-type FPE where $\alpha=1$. However, this does not hold in higher dimensions, where A-type FPE is no longer identical to any $\alpha$-type FPE in general, even when $\B$ is constant. This is because the anti-symmetric matrix $\Q$ can also give the additional fixing term. While when $\B$ is constant, all the $\alpha$-type integration methods are identical to each other. Following two examples serve to show the differences between I-type and A-type processes.

\begin{ex:ex}{(Shift of stable point)} \label{ex:1}
    \begin{figure}
      \includegraphics[scale = 0.5]{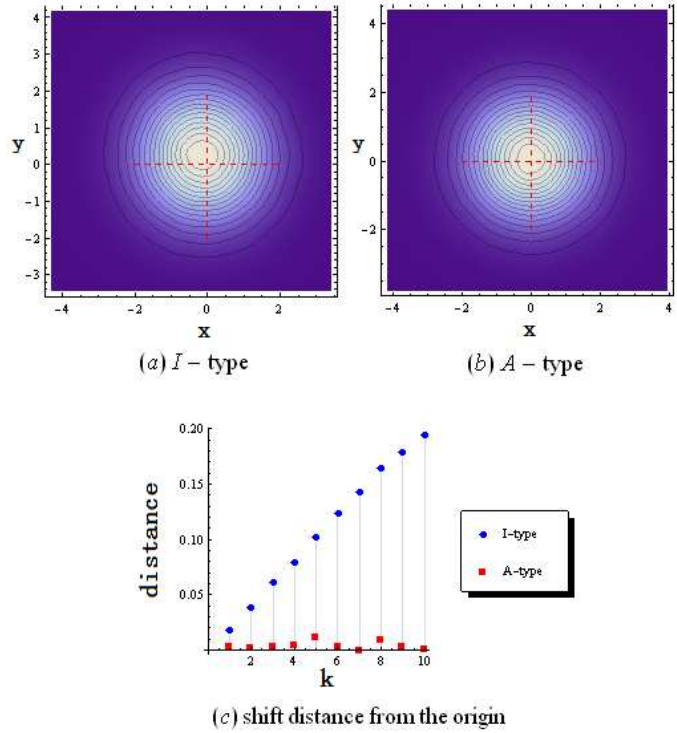}
    \caption{(Color online) Top: Density histograms of long time trajectories of I-type process(a) and A-type process(b) in Example \ref{ex:1}. The origin is pointed out with intersection of red lines. Different colors indicate different density, and lighter color represents higher density. Down: The distance between the origin and the most probable state(the lightest point in (a)(b)) for I-type process(Blue) and A-type process(Red) with different values of the parameter $k$.}
    \label{fig:1}       
    \end{figure}
Figure \ref{fig:1} (a)(b) shows the density histograms of long time trajectories of I-type process and A-type process. These two processes are generated from an identical SDE with constant $\D$ and state dependent $\Q$ and the drift force of the SDE has the origin as its stable fixed point. For the I-type process, there is a shift between its most probable state and the origin, while A-type process keeps the origin as its most probable state. Figure \ref{fig:1} (c) shows the sampling shift distance from the origin via changing a parameter of $\Q$. Detailed parameters can be found in appendix.
\end{ex:ex}

\begin{ex:ex} \label{ex:2} (Change of global shape)
    \begin{figure}
      \includegraphics[scale = 0.5]{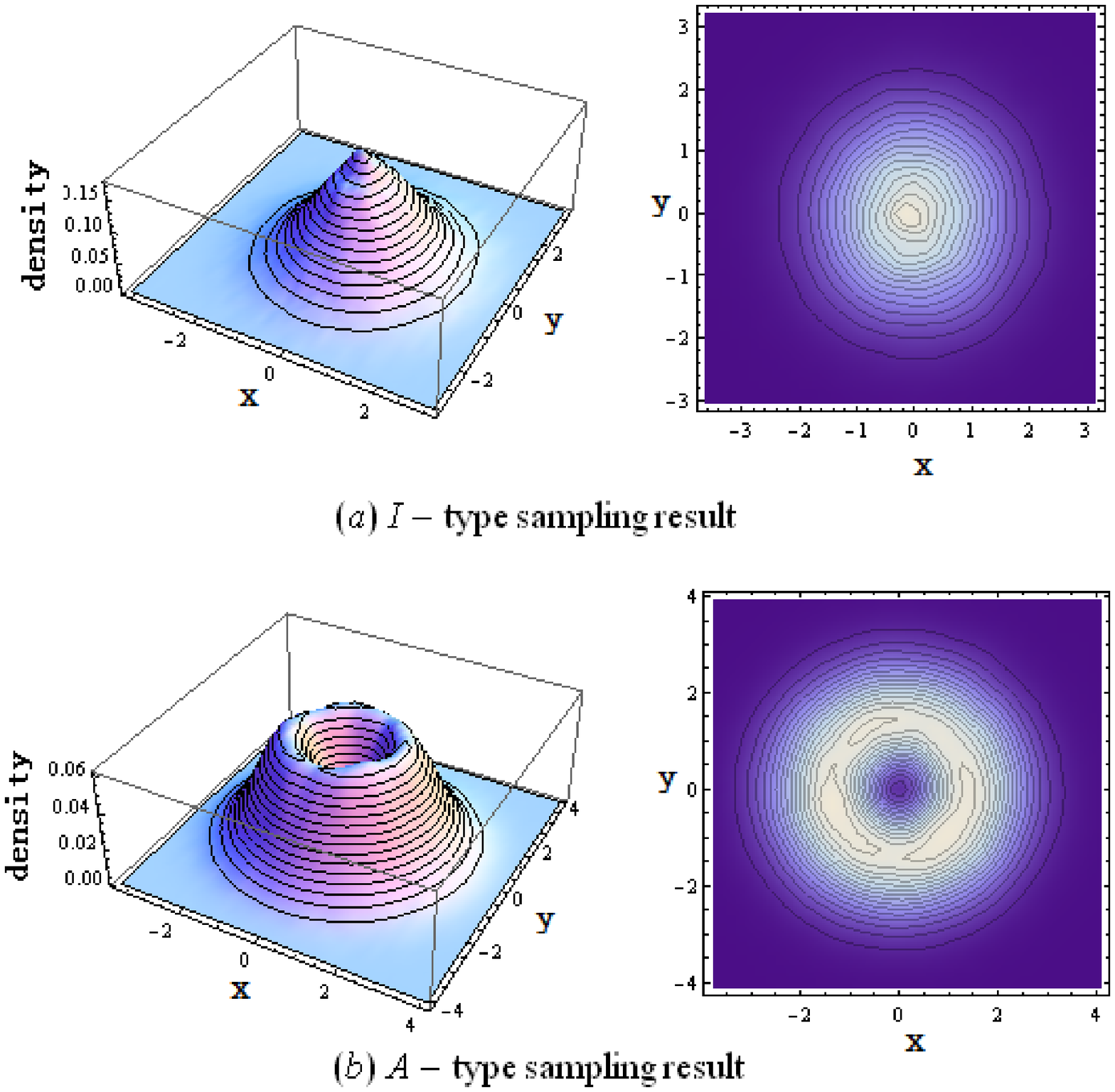}
    \caption{(Color online) Density histograms of long time trajectories of I-type process(Top) and A-type process(Down) in Example \ref{ex:2}. Left panels show the density difference with altitude and right panels with color lightness.}
    \label{fig:2}       
    \end{figure}
Figure \ref{fig:2} shows the density histograms of long time trajectories of I-type process and A-type process. These two processes are generated from an identical SDE with state dependent $\D$ and zero $\Q$. Note that the drift force of the SDE has a circle as its stable fixed points and the origin as its unstable fixed point. In this setting, the stationary distributions of I-type and A-type processes show significant differences. For the I-type process, the unstable fixed point becomes the most probable state. On the contrary, A-type process keeps the stable fixed points (the circle) as its most probable states.
\end{ex:ex}

Two remarks are in order. First, the noise induced shifts of bifurcations have been amply studies in literature, see for example, \cite{arnold_2003}, but here we study it from A-type approach perspective. Second, we should point out that those examples may well be what may occur experimentally. A real physical experiment is discussed in next section as a concrete example to further illustrate such possibility.
\subsection{Stationary distribution- 1 dimensional case}
When dealing with a real process and the SDE is given, how to choose the integration method? Since the FPE uniquely determines the process, we can refer to its stationary distribution. As we discussed in the previous section, different interpretations will result in different stochastic process, thus lead to different stationary distributions. A proper interpretation should lead to the stationary distribution of the real process. In the following, we will consider stationary distribution of different methods in one dimensional case. The temperature $\theta$ is set to 1 for simplicity.

For the following one dimensional SDE
\begin{equation}
\d x = -D(x)\phi'(x)+ \B(x)\d\W(t),
\end{equation}
the corresponding I-type FPE is
\begin{equation}
\partial_t \rho(x)= -\partial_{x}[-D(x)\phi'(x)\rho(x)-\partial_{x}(D(x)\rho(x))].
\end{equation}
Applying Theorem \ref{thm:ito alpha relation}, note that only in one dimensional case, once $D$ is determined, different possible forms of $\B$ will not affect the modification term. Hence we can get the $\alpha$-type FPE as following
\begin{equation}
\partial_t \rho(x)= -\partial_{x}[-D(x)\phi'(x)\rho(x)-D^{\alpha}(x)\partial_{x}(D^{1-\alpha}(x)\rho(x))].
\end{equation}
Its stationary distribution is
\begin{equation}
\rho(x)\propto D^{\alpha-1}(x) e^{-\phi(x)}.
\end{equation}
Note that when $\alpha=1$, the above stationary distribution is Boltzmann-Gibbs distribution, same with A-type interpretation. A real physical experiment done recently favors this interpretation in one dimensional case \cite{volpe_2010}.

Above results show that even for one dimensional case, when $D$ is not constant, different integration methods lead to different stationary distributions. This fact reminds us to pay attention when choosing integration methods.

For higher dimensional cases, usually one can not write down the stationary distribution for $\alpha$-type integration directly, due to the emergence of the breaking detailed balance term $\T$. While for A-type interpretation, one can always write down the stationary distribution directly, which is the Boltzmann-Gibbs distribution (Equation \ref{eq:bgd}).

\section{Conclusion}
In this paper, we discussed the relations between several SDE integration methods, including classical ones and a recent framework introduced by one of us. Their experimental implications are also discussed.

The richness nature of SDE has long been known by literature. Different integration methods are all mathematically consistent within their own framework. Thus there may be illusion that we can apply them arbitrarily as we like. While the fact is we should pay attention to the pseudo freedom, since the real process is unique. In real process it is always hard to determine which method we should take, but we can always refer to the FPE for help, since it uniquely determines a process.

A-type framework is mysterious at first sight. Instead of specifying integration explicitly, it directly connecting SDE and FPE in their special form. With the special form of FPE, its stationary distribution enjoys the property of Boltzmann-Gibbs distribution. Though it does not specify how the actual process is integrated, by comparing the FPE with I-type FPE we can find a way to convert A-type parameters into I-type ones. Hence simulate A-type processes by I-type processes. It is interesting to note that A-type decomposition has a discrete state counterpart, which will be discussed elsewhere.

We have shown that A-type SDE is not a special kind of $\alpha$-type SDE. So $\alpha$-type has not cover all the existing interpretations of SDE. We now end here with a question for readers: is there a more general form of stochastic integration which can unify all the mentioned types of integration?

\begin{acknowledgements}
The authors would like to express their sincere gratitude for the helpful discussions with Song Xu, Xinan Wang, Yian Ma, Ying Tang.
This work was supported in part by the National 973 Project No.~2010CB529200 and Project No.91029738(P.A.); and by the Chinese Natural Science Foundation No.~NFSC61073087 (R.Y., J.S. and B.Y.).
\end{acknowledgements}

\bibliographystyle{plain}
\bibliography{sde}

\appendix
\section{Proof of theorem \ref{thm:ito alpha relation}}
\begin{prf:prf}
We only need to consider the difference between $\alpha$-type and I-type processes ($\alpha=0$),
the difference in drift part is given by
\begin{equation}
  [\f(\x(t+\alpha \d t) ) - \f(\x(t))] \d t = \mathrm{O}\left( \x(t+\alpha \d t) - \x(t) \right)\d t = \mathrm{O}( \d t^{1.5}) = \mathrm{o}(\d t).
\end{equation}
This means the difference in drift part is negligible. The difference in diffusion part of
$i$th coordinate is given by
\begin{align}
 & \sum_{j}[\B_{ij}(\x(t+\alpha \d t))-\B_{ij}(\x(t))]\d \W_{j}(t) \\
\label{eq:pf1}=& \sum_j [\B_{ij}\left(\x(t)+ \B(\x(t))\d \W(\alpha t) \right)-\B_{ij}(\x(t))] \d \W_{j}(t) \\
\label{eq:pf2}=& \sum_{j} \sum_{k,s} (\partial_{k} \B_{ij}(\x)) \B_{ks}(\x)\d \W_{s}( \alpha t) \d \W_{j}(t) + \mathrm{o}(\d t)\\
\label{eq:pf3}=& \alpha \sum_{j} \sum_{k} (\partial_{k} \B_{ij}(\x)) \B_{kj}(\x)\d t + \mathrm{o}(\d t).
\end{align}
We can use interpretation of SDE over time interval $t$ to $t+\alpha \d t$ to get Equation \ref{eq:pf1}. Here $\d\W(\alpha t)$ is
a short notation for change of Wiener Process over time $\alpha \d t$. Equation \ref{eq:pf2} is given by first order expansion. We get the last equality using the following facts
\begin{equation}
\d\W_{i}(\alpha t)\d \W_{j}(t) =\left\{
  \begin{array}{ll}
  \R_1(t) & i\neq j, \\
  \alpha \d t  + \R_2(t)  & i = j.\\
  \end{array}
  \right.
\end{equation}
Here $\R_1(t)$ and $\R_2(t)$ are zero mean noise with standard deviation of order $\mathrm{o}(\sqrt{\d t})$.
These small noise will not harm the result and can be ignored according to the following theorem.
\begin{thm:thm}
  The zero mean noise in $\d \textbf{x}$ with standard deviation of $\mathrm{o}(\sqrt{\d t})$ can be ignored without influencing
  the result of stochastic integration.
\end{thm:thm}
\begin{prf:prf}
  Let us denote the noise term $\R(t)$. Consider the stochastic integration of the noise term over a time interval
  \begin{equation}
    \X = \int_{t=0}^{T} \R(t) = \lim_{N\rightarrow \infty}\sum_{i=1}^{N-1}\R(t_i)
  \end{equation}
  We can find that $\X$ is a random variable with zero mean and zero variance (due to the fact that variance of $ \R(t)$ is $\mathrm{o}(\d t)$ ).
  This means $\X$ goes to 0 by mean-square limit.
\end{prf:prf}
\end{prf:prf}

\section{Detailed parameters of examples}
\subsection{Example \ref{ex:1}}
The SDE for the I-type process is
\begin{equation}\label{eq:example1}
\d\x = -(\D+\Q)\nabla \phi \d t + \B \d\W(t), \B\B^\tau=2\D
\end{equation}
Where
\begin{equation}
\x = \left(
      \begin{array}{c}
        x \\
        y \\
      \end{array}
    \right),
\B = \left(
      \begin{array}{cc}
        10 & 0 \\
        0 & 10 \\
      \end{array}
    \right),
\Q = \left(
      \begin{array}{cc}
        0 & k y \\
        -k y & 0 \\
      \end{array}
    \right),
\phi  = (x^2+y^2)/2
\end{equation}
Where $k$ is an integer vary from 1 to 10. The A-type process for Equation \ref{eq:example1} is the following I-type process:
\begin{equation}
\d\x = [-(\D+\Q)\nabla \phi+ \Delta \f] \d t +\B \d\W(t)
\end{equation}
where
\begin{equation}
\Delta \f= \left(
      \begin{array}{c}
         k  \\
        0 \\
      \end{array}
    \right)
\end{equation}

\subsection{Example \ref{ex:2}}
The SDE for the I-type process is
\begin{equation}\label{eq:example2}
\d\x = - \D \nabla \phi \d t + \B \d\W(t),\B\B^\tau=2\D
\end{equation}
Where
\begin{equation}
\x = \left(
      \begin{array}{c}
        x\\
        y \\
      \end{array}
    \right),
\B = \left(
      \begin{array}{cc}
        \sqrt{x^2+y^2} & 0 \\
        0 & \sqrt{x^2+y^2}\\
      \end{array}
    \right),
\phi  = -\log{(x^2+y^2)} + (x^2+y^2)/2
\end{equation}
It is easy to see that the above I-type process is the equivalent I-type process for the following A-type process with same $\B$ and different $\phi$:
\begin{equation}
\d\x = - \D \nabla \phi' \d t + \B * \d\W(t),\B\B^\tau=2\D
\end{equation}
Where
\begin{equation}
\phi'  = (x^2+y^2)/2
\end{equation}
So its stationary distribution is
\begin{equation}
\rho\propto \exp\left[-(x^2+y^2)/2\right]
\end{equation}
This stationary distribution has only one minimum point at origin.

The A-type process for Equation \ref{eq:example2} is the following I-type process:
\begin{equation}
\d\x = [-\D\nabla \phi+ \Delta \f] \d t + \B \d\W(t)
\end{equation}
where
\begin{equation}
\Delta \f= \left(
      \begin{array}{c}
        x  \\
        y \\
      \end{array}
    \right)
\end{equation}
And its stationary distribution is
\begin{equation}
\rho\propto \exp\left[ \log(x^2+y^2)-(x^2+y^2)/2\right]
\end{equation}

\end{document}